\documentclass{article}

\PassOptionsToPackage{square,sort&compress,comma,numbers}{natbib}

\usepackage[final]{neurips_2020}

\usepackage[utf8]{inputenc} 
\usepackage[T1]{fontenc}    
\usepackage{hyperref}       
\usepackage{url}            
\usepackage{booktabs}       
\usepackage{amsfonts}       
\usepackage{nicefrac}       
\usepackage{microtype}      

\title{Continuous Subject-in-the-Loop Integration: Centering AI on Marginalized Communities}

\author{%
  Francois Roewer-Despres \\
  Department of Computer Science \\
  University of Toronto \\
  \texttt{francoisrd@cs.toronto.edu} \\
  \And
  Janelle Berscheid \\
  Department of Computer Science \\
  University of Saskatchewan \\
  \texttt{janelle.berscheid@usask.ca} \\
}

\begin{document}

\maketitle

\begin{abstract}
Despite its utopian promises as a disruptive equalizer, AI---like most tools deployed under the guise of neutrality---has tended to simply reinforce existing social structures. To counter this trend, radical AI calls for centering on the marginalized. We argue that gaps in key infrastructure are preventing the widespread adoption of radical AI, and propose a guiding principle for both identifying these infrastructure gaps and evaluating whether proposals for new infrastructure effectively center marginalized voices.
\end{abstract}

\section{Introduction}

Neutrality as an impartial, apolitical stance is an illusion. Instead, neutrality implicitly endorses the status quo~\cite{whittaker_ai_2018, hoffmann_where_2019, green_good_2019, zajko_conservative_2020}. It is a privilege afforded to those who benefit from existing power structures. As such, it is not surprising that ``neutral'' AI deployment in our world has reinforced existing social structures, injustices, and disadvantages~\cite{zajko_conservative_2020, birhane_algorithmic_2019, benthall_racial_2019, oneil_weapons_2016, noble_algorithms_2018, benjamin_race_2019, onuoha_peoples_2018, yeshi_abolish_2019}. These systematic disadvantages exist ``to produce systematic advantages for others''~\cite{zajko_conservative_2020}; like many technologies, AI tends to catalyze the concentration of power in the hands of the few at the expense of the vulnerable~\cite{zajko_conservative_2020, birhane_algorithmic_2019, keyes_counting_2019, mcquillan_ai_2020, cifor_feminist_2019}. For example, AI makes many of the most exploitative practices of capitalism more efficient~\cite{zajko_conservative_2020, oneil_weapons_2016, keyes_counting_2019}.

Given this worrying trend, many scientists and activists have called for debiasing AI---being more aggressively neutral to attempt true neutrality. Unfortunately, these technical solutions, including attempts at gathering unbiased data and at ensuring algorithmic blindness to existing data bias, fall short due to both technical and non-technical limitations~\cite{zajko_conservative_2020, birhane_algorithmic_2019, benthall_racial_2019, cifor_feminist_2019, keyes_counting_2019, friedler_impossibility_2016, buolamwini_gender_2018, raji_actionable_2019, polack_beyond_2020, bowyer_criminality_2020}. Fundamentally, the problem is that `removing bias' frames the issue superficially and negatively, revealing what is undesirable without suggesting thoughtful alternatives~\cite{green_good_2019, zajko_conservative_2020}. After all, ``a bias-free world could be one where every person is equally surveilled and controlled''~\cite{zajko_conservative_2020}. Instead, many scholars urge us to think positively about the world we want to create, and to reimagine AI's role within it~\cite{green_good_2019, zajko_conservative_2020, birhane_algorithmic_2019, benthall_racial_2019, onuoha_peoples_2018, yeshi_abolish_2019, keyes_counting_2019, mcquillan_ai_2020, mcquillan_towards_2019, costanza_design_2020}.

However, this positive reimagining is limited when marginalized voices are ignored in favour of those already in power. Advantaged voices tend to envision a world where their advantage remains~\cite{green_good_2019, zajko_conservative_2020, birhane_algorithmic_2019, giridharadas_winners_2019, farrell_prodigal_2020}, seeking ``to do more good, but never less harm''~\cite{giridharadas_winners_2019}. Radical approaches to AI call for explicitly centering marginalized voices in these discussions~\cite{zajko_conservative_2020, birhane_algorithmic_2019, onuoha_peoples_2018, yeshi_abolish_2019, cifor_feminist_2019, keyes_counting_2019, mcquillan_ai_2020, mcquillan_towards_2019, costanza_design_2020}, allowing us to imagine new ways of thinking about AI that depart from the status quo.

Unfortunately, the concept of radical AI is often met with skepticism and resistance~\cite{zajko_conservative_2020}. Radical change involves long timescales, abstract ideas, and confronting difficult problems. In contrast, incremental change is easier to envision and implement~\cite{green_good_2019}, and ``remains entrenched through longstanding proximity to powerful interests''~\cite{zajko_conservative_2020}. Given this pushback from entrenched powers, how do we ensure marginalized voices are not suppressed, despite our best efforts? We argue that critical infrastructure can meaningfully counterbalance these dominant structures.

\section{On Infrastructure}

Counterbalancing power requires not only creativity, will, and vision~\cite{giridharadas_winners_2019}, but also access to resources, expertise, and partnerships. That is, the process of radical change requires a foundation of critical infrastructure. Infrastructure consists of physical assets that increase productivity (economic infrastructure, like roads and power grids), public service assets that improve lives (social infrastructure, like healthcare and green spaces), and institutional assets that help maintain society (soft infrastructure, like laws and education)~\cite{vaughan_strategic_2012}. We frame barriers to the widespread adoption of radical AI as gaps in critical infrastructure.

This framing has several benefits. First, it provides a novel perspective on the problem of ``neutral'' AI: without additional infrastructural constraints, AI systems simply adhere to existing societal structures, replicating their injustices. When no infrastructure exists to prevent it, private companies face few consequences for inflicting harm (intentional or otherwise) on marginalized communities. AI systems can cause considerable harm, such as by unjustly denying loans, miscalculating recidivism likelihood, or failing to detect cancer. However, they lack the careful regulation of other industries known to have the potential for harm. For example, new drugs developed by the pharmaceutical industry cannot be distributed without ensuring they are safe and effective, but new AI systems generally do not face similar restrictions. This situation cannot be fixed on a case-by-case basis; it requires infrastructure---such as regulatory bodies, legal access and accountability, public educational resources, and community partnerships---to ensure potential harms can no longer be ignored.

Second, filling these infrastructure gaps creates assets designed to oppose systemic forces, while still operating within those systems. As such, this approach can be implemented with less friction than more radical changes, yet creates a strong foundation with which to implement such changes in the future. In the meantime, it provides a stabilizing ``counterpower''~\cite{keyes_counting_2019} that prevents further erosion of freedoms for marginalized groups.

Lastly, robust public infrastructure emphasizes the common good. Relying solely on private actors to solve AI's injustices leads to ``solutions'' that are piecemeal, inconsistent, or even harmful. Public infrastructure has a stronger focus on consistency, equality of access, and equality of participation, allowing communities to work together towards a shared goal. For example, a company's internal ethics board may be pressured to put the interests of the company ahead of its stated goal of ethical oversight. Furthermore, different companies' ethics boards may not align in their ethical principles or evaluation criteria. A community-focused initiative, such as a people's council~\cite{mcquillan_ai_2020}, provides more equitable and consistent infrastructure for exploring the impacts of AI systems.

From this perspective, the need for infrastructure that fills these gaps becomes clear. However, identifying which specific infrastructure gaps exist is challenging. In addition, many infrastructure proposals aimed at filling these gaps may appear to center the marginalized without truly doing so. We thus propose a guiding principle to facilitate these identification and evaluation processes.

\section{A Guiding Principle}

The potential for AI to cause harm is a social problem, yet AI is fundamentally technological, so this potential for harm is equally a technical problem. Many attempts at preventing harm fall short precisely because they fail to address this dual nature. However, it is also insufficient to propose solutions that address both facets of the problem independently. Rather, we argue that the practice of evaluating social harms must be tightly integrated into the technical development process.

We propose a guiding principle for centering the marginalized, \underline{C}ontinuous \underline{S}ubject-in-the-\underline{L}oop \underline{I}ntegration (CSLI), that fulfils this requirement. CSLI is inspired by the software engineering practice of continuous integration\footnote{Here, we focus on the high-level aspects of CSLI as a guiding principle, since, given its pragmatic inspiration, we hope that its ability to guide the low-level implementation of software engineering processes is self-evident.}, where developers frequently integrate their individual work to ensure proper alignment across a team. Compared with doing so only at the end of a project, this measure of quality control significantly reduces development time and errors. X-in-the-loop integrates X into the iterative cycle (``the loop'') in software development. In this case, X refers to subjects---the (often marginalized) targets of AI systems. ``Subject'' is a deliberate word choice: since many AI systems are not properly vetted before deployment, their targets often double as experimental test subjects.

As a guiding principle, CSLI promotes infrastructure that tightly integrates subjects into the development of AI systems to catch and fix problems early, rather than allowing harmful errors and assumptions to accumulate. CSLI mandates continuous subject involvement in the development process---from requirements gathering, through development, and continuing into deployment, including all future updates and changes until the system is retired. This ensures subjects are always informed of the state of software targeting them, and gives them a persistent voice throughout. Early integration also ensures that ethics are considered from the start, rather than merely as an afterthought.

However, we emphasize that CSLI is not just about negatively constraining developers and harmful applications. Radical AI involves positively reimagining AI's role in the world we want to create, and positive CSLI-guided infrastructure facilitates this process. For example, positive infrastructure could enable community networks to be paired with a team of developers to help realize their vision; or perhaps AI systems, with the right infrastructure, can help achieve the United Nations' Sustainable Development Goals more quickly and at scale, in ways that meaningfully empower communities~\cite{elsayed_response_2019}.

CSLI can be applied in two ways. First, it can reveal critical infrastructure gaps in the existing system. Second, it can evaluate whether a proposal to fill these gaps achieves its aims. Although CSLI can be applied directly in both cases, three criteria naturally emerge to help structure this process:

\paragraph{Coverage.} What ensures all impacted subjects are consulted? Developers cannot be selective in their community involvement. In addition, which infrastructure types (economic, social, and soft) are being considered? Proposals need not cover all types, but awareness of their limited scope is critical.

\paragraph{Integration.} How tightly are subjects integrated into the development process? How forefronted are their concerns? Are they genuinely interested in the project? From day one, subjects should be proactively (tightly, and before development) rather than reactively (loosely, and after deployment) integrated. In addition, what provisions establish and strengthen partnerships within and between communities, to facilitate the sharing of insights and concerns across wider groups?

\paragraph{Outlook.} Are the positive, forward-oriented goals of radical AI promoted? Negative infrastructure is not bad---in fact, it is very much needed---but positive infrastructure should not be forgotten simply because it is harder to envision and implement. In addition, are subjects able to propose their positive visions for reimagining AI? What ensures these visions are realized?

Despite concerted effort in recent years to address problems with AI (e.g.~\cite{brundage_toward_2020, berscheid_beyond_2019}), infrastructure gaps clearly remain. Using CSLI, we highlight a few specific gaps in the North American context:
\begin{itemize}
    \item \textbf{Regulatory infrastructure.} There is no regulatory system mandating that developers meaningfully involve targeted communities, and no clear consequences for failing to consult subjects or take community feedback in account. Here we find either a lack of coverage (if not all subjects are being consulted) or a lack of integration (if the concerns of consulted subjects are being ignored). This makes ethics-washing and empty gestures very easy~\cite{metcalf_owning_2019}.
    \item \textbf{Legal infrastructure.} There are no provisions for closely involving subjects in audits, or for non-excludable recourse when wronged. This indicates a lack of integration, as even in these post-deployment scenarios, subject involvement is necessary to ensure continuing developer accountability.
    \item \textbf{Support infrastructure.} There is no network allowing advocacy groups, independent ethics boards, and other communities to create partnerships to aid each other with consultations, and no public recordkeeping to share insights and precedents. This lack of integration across wider groups increases both the labour requested of marginalized communities and the risk that developers overlook relevant concerns identified elsewhere. Additionally, this indicates a lack of outlook, as the ability for communities to realize their positive visions of AI hinges on their ability to communicate and organize effectively.
    \item \textbf{Educational infrastructure.} There are no common, publicly-accessible educational resources for communities. This is an outlook problem: while it is unreasonable to expect all subjects to become AI experts, creating radical change requires that at least a high-level awareness and understanding of AI be ubiquitous within communities.
\end{itemize}

For both support and education, a collective body of knowledge both reduces the burden on individual experts in a given community, and lessens the impact of a lack of experts within any one community.

\subsection{Case Study}

Using CSLI as a guiding principle, we reflexively~\cite{green_good_2019} examine our own previous work on developing an actionable accountability framework for AI systems~\cite{berscheid_beyond_2019}. This framework aims to regulate AI systems in critical domains with significant impact on quality of life (e.g.~livelihood, health, or freedom). It distinguishes between three parties: subjects and developers, as well as clients (for example, a hospital) who license AI systems from developers to deploy them in some context.

Developers and clients have a many-to-many relationship: developers' systems may be licensed to multiple clients, and a single client can license systems from multiple developers. This creates tension between developers and clients: they tend to blame each other when subjects are wronged. In such cases, the framework's two-stage structure helps determine accountability through concrete validation, disclosure, and appeal processes. For each licensed system, each party files a separate document delineating its responsibilities. The developer's document details technical responsibilities, including validation processes such as intended use cases and evaluation metrics. Once approved by a qualified third party (analogously to patent applications), clients file their own document detailing deployment responsibilities, including disclosure and appeal processes.

CSLI's three evaluation criteria reveal the benefits of this framework, while also making its shortcomings obvious. In terms of coverage, the framework fills some key regulatory and legal (i.e. soft) infrastructure gaps, and its disclosure and appeal processes are steps towards including impacted subjects in at least some part of the system. However, while the framework does provide a mechanism (validated scope) for identifying relevant subject communities, leaving this process up to developers and clients creates an opportunity for them to be selective in their community involvement, and thus does not go far enough towards ensuring all impacted subjects are included.

In terms of integration, this framework prioritizes a developer-client relationship that improves accountability and increases awareness of AI's potential to harm subjects. However, this only guarantees that they have carefully considered issues concerning subjects, not that subjects themselves were actually consulted. In addition, the framework neither incorporates subjects in the development process, nor allows them to question whether AI should be used at all (as not all problems need technical solutions~\cite{green_good_2019, mcquillan_ai_2020, costanza_design_2020}). Instead, it only mandates that subjects can audit or challenge AI-made decisions after deployment. These are reactive rather than proactive approaches that fail to appropriately integrate subjects.

In terms of outlook, rather than reimagining how AI can change the world for the better, this framework takes a more punitive approach. 
Subjects are only integrated into the process to monitor, surveil, and approve or reject systems. It allows for, but in no way encourages subjects to think positively about using AI to achieve their community goals. By emphasizing the control of bad AI, rather than imagining and creating good AI, this framework only promotes negative infrastructure.

\section{Discussion}

Despite its potential as an equalizer, AI tends to simply reinforce existing social injustices. Unfortunately, debiasing AI is not a viable solution to this problem. Rather than thinking negatively in these terms, we need to positively reimagine AI's role in the world we want to create. Inclusive reimagining requires centering the discussion on the marginalized, rather than the advantaged. However, this centering is not easily achieved, since existing systemic forces will resist it at every turn. We framed this barrier as infrastructure gaps, arguing that the process of radical change requires a foundation of critical, counterbalancing infrastructure. However, many proposals attempting to fill these gaps fail to properly center the marginalized. We thus introduced CSLI as a guiding principle for identifying infrastructure gaps and evaluating the efficacy of proposals aimed at filling them, using three criteria---coverage, integration, and outlook---to ensure marginalized voices are explicitly centered.

Crucially, CSLI addresses both the social and technical problems around AI. In this way, CSLI helps fulfill the promise of radical AI: not just fighting against existing, oppressive technologies, but also dreaming up AI systems that benefit all. As we have shown, this dream is not easily accomplished. And yet, by focusing on centering marginalized voices as the driving tenet of potential AI futures, and by framing the process in terms of communities and infrastructure, rather than the efforts of a few individuals and companies, we can build strong foundations for a more positive AI ecosystem.

\end{document}